# Atomic Dipole Traps with Amplified Spontaneous Emission: A Proposal

Jean-François Clément, Véronique Zehnlé, Jean Claude Garreau, and Pascal Szriftgiser[*]

**Abstract:** We propose what we believe to be a novel type of optical source for ultra-cold atomic Far Off-Resonance optical-dipole Traps (FORTs). The source is based on an Erbium Amplified Spontaneous Emission (ASE) source that seeds a high power Erbium Doped Fiber Amplifier (EDFA). The main interest of this source is its very low coherence length, thus allowing an incoherent superposition of several trapping beams without any optical interference. The behavior of the superimposed beams is then a scalar sum greatly simplifying complex configurations. As an illustration, we report an estimation of the intensity noise of this source and an estimation of the atomic excess heating rate for an evaporative cooling experiment application. They are both found to be suitable for cold atoms experiments.

Laboratoire PhLAM, UMR CNRS 8523, IRCICA FR CNRS 3024, Université Lille 1, 59655 Villeneuve d'Ascq Cedex, France

[*]Corresponding author. E-mail address: Pascal.Szriftgiser@univ-lille1.fr

PACS: 37.10.Gh Atom traps and guides

Keywords: Amplified spontaneous emission, cold atoms, optical traps, heating rate

## 1. Introduction

Far off-resonance optical-dipole traps (FORTs) are at the heart of a constantly increasing number of cold atom experiments [1-3]. They are widely used as a convenient way to manipulate cold atoms or molecules while minimizing decoherence processes such as spontaneous emission. As a very incomplete list of examples, this includes: atomic interferometers, evaporative cooling to produce Bose-Einstein Condensates (BECs), time dynamics and momentum exchange experiments, [4-8]… In the present paper we focus on the possibility of improving the all-optical BEC evaporative cooling setup with trapping beams in the optical telecommunication window, around 1550 nm. Indeed, roughly two general categories of experiments can be made. Firstly, the user needs to generate complex potentials such as a grating in one, two or three dimensions. In this case, the laser that generates the potential must have a high coherence level to preserve a good interference contrast at the crossing of the beams. This is generally not a problem. Nowadays lasers currently have linewidths of the order of the MHz, or below. These linewidths corresponds to a coherence length of a few hundreds of meters or better; a value that will be enough, by far, for most applications. On the other hand, one might be interested by the potential resulting from of the scalar sum of several laser beam intensity. The interferometric feature of light is then unwanted. This is generally the case in the preliminary phases of the experiment where the atomic cloud needs to be hold for an additional preparation procedure as, e.g., a transfer from one trap to another, or for a further cooling stage. The question is then how to get rid of light interferences? The first idea is to use as many laser with different frequencies as the number of the required beams. If the frequency difference between each laser is greater than the characteristic frequencies of the atomic system, their beat note will not affect this system. However, this solution is both complex and expensive compared to a single laser source realization. The second possibility is to use a single laser source, and to cross at right angle the polarizations of its multiple beams in the overlapping area. If this is feasible, this solution has several drawbacks. The number of available beams is typically limited to two or three, but especially, it is very difficult to check that the polarizations have the required orientation in an ultra-high vacuum cell. This is even more difficult, if not impossible, with high numerical aperture optics and short Rayleigh lengths used, for instance, in single atom experiments [9].

The simplest solution is probably to employ a spectrally broad light source with a very short coherence length, so that a small difference in the optical paths will be enough to insure a complete loss of coherence. This solution is especially attractive since there is a very intense activity in the field of both fiber lasers and amplifiers. These devices are widely available with all the advantages of fiber devices such as: low cost, low weight, high compactness, insensitivity to acoustic noise, easy beam delivery… Furthermore, they can provide very high power with a continuous operating mode. For instance, up to 100 W are presently available with Erbium Doped Fiber Amplifiers (EDFA) in the optical telecommunication C-band (from ~1530 to ~1570 nm). The C-band is of particular interest. Indeed, an optical-dipole trap to achieve Bose-Einstein condensation has recently been demonstrated with an Erbium doped fiber laser [6]. This recent work showed demonstrated the feasibility of an efficient production of a $^{87}$Rb BEC in a 1565 nm optical-dipole trap. In this work, tThe loading stage of the experimental sequence needs a two-beam configuration. The with two beams crossing each other at roughly right angle, and one of them is frequency shifted with respect to the other one with by an acousto-optical modulator.

However, this configuration is not optimal. In fact, the key point of the loading efficiency is based on the common trap volume of the beams which could be improved if these two beams were aligned. With an ASE-based FORT, one could propose a two-beam geometry based on a single retro-reflected beam without any acousto-optical modulator. A first wide beam is then dedicated to the loading process, and can be re-circulated to provide a tightly confining beam confinement, making possible an efficient evaporative cooling sequence. The ASE-based FORT is notably very well suited for this multi-beam configuration without paying attention to interference effects and optical power losses, i.e. using wave plates for each beam or frequency shifting one of them. It would thus allows to create FORTs for almost all the alkali-metal atoms used in cold-atoms experiments (Li, Na, K, Rb and Cs) with only one powerful laser source, this with a high efficient loading rate and an improved evaporation stage for many different beam-geometries: dimple, Misaligned Crossed-Beam FORT [6, 10]…

In the C-band, both fiber lasers and amplifiers are available. High power fiber lasers have spectrally broad characteristics capabilities. Typically, when they are not optimized to be spectrally narrow, their width ranges from a few tens of GHz to a few hundred [11]. At first sight, they could be good candidates for our present requirements. Nevertheless, they are highly multimode. Their spectrum is composed by a multitude of longitudinal modes with a frequency spacing determined by their free spectral range. Individually, each of these modes is spectrally narrow (~1 MHz). These individual modes have a very long coherence length. Although very attractive, this kind of lasers cannot be used as an incoherent source. To both overcome this difficulty and take advantage of the optical telecommunication technology, we propose to construct a spectrally broad optical source starting from the spontaneous emission generated by an erbium doped fiber. These sources are usually pumped at 980 nm by a multimode laser diode. They are commercially available, but can also easily be implemented. They might also simply be obtained from a low power EDFA without seeding it. Amplified Spontaneous Emission (ASE) sources have a very broad spectral width. They roughly cover the whole C-band (~3 THz). This corresponds to a very short 1 μm coherence length, an ideal value for our present requirement. Furthermore, one should notice that an ASE source is not a laser, a critical feature. Indeed, in a general manner, the problem with broad multimode lasers – including for instance high-power multimode laser diodes – is not restricted to the high coherence level of their individual modes. Notably, because of Kerr effect, four wave-mixing, hole-burning, each of these mode strongly interacts with each other. This leads to a complex chaotic dynamic of their spectrum, rendering them unusable for cold particles handling. As mentioned above, an ASE source is not a laser. It displays a continuous spectrum with a spectral density too low to induce observable non linear effects. These sources are thus very stable (they are noisy, but their spectra are stable).

Cold atom experiments need very low intensity noise light. Intensity noise will induce additional heating rate and possible excess atom losses [12]. As an ASE source is a 100% noisy light source, this proposition could appear paradoxical at first sight. However, the noise of an erbium ASE source is distributed over a very wide frequency range (~3 THz), the fraction of its spectrum overlapping the atomic system resonances (in the kHz range) is very low. In the following, we show that for well chosen parameters, this source can be seen as a low noise source with features comparable to common lasers used in every day cold atom laboratories.

In section 2, a possible setup is presented and discussed. An experimental estimation of the noise level for a low power ASE source is also reported. Section 3 is devoted to a numerical estimation of the heating rate with parameters adapted for two practical applications: evaporative cooling of Rubidium atoms, and single atoms trapping.

## 2. Experimental setup

Before discussing the ASE source, it is interesting to mention that broad band sources have been successfully used or proposed for light-matter interaction experiments. For instance, white light is of potential interest for optical tweezers application for cold atoms [13], or small biological items [14]. The proposition to use an ASE source is new to the best of our knowledge. A scheme for a realization is proposed figure 1. The starting point is a commercial low power Erbium ASE source. The simplest configuration is to directly amplify this source with a high power EDFA. Typically, 1 mW is sufficient to seed a high power EDFA, a value easily reach by most ASE sources. The grey insets indicate possible optional components to adapt the device for particular requirements. The use of a high power EDFA is a very convenient way to obtain a high power source with potential complex features. It is simpler to start by building a low power light source with the desired features, then, amplify it.

The first proposed option is the polarization beam splitter (PBS). In its basic configuration, an ASE source is not polarized. It is composed of a statistics mixture of the two polarization states without any correlations between them. This is a very convenient feature, and one can take advantage of it. As the source is depolarized, there is no need to perform any polarization alignment. This is not only potentially the case for the final dipole trap beams, but also for the fiber part of the device. This implied that the fibered system will be stable even if realized without polarization maintaining (PM) fibers and components; an interesting point as the cost a non PM EDFA is greatly reduced compared to a PM one. On the other hand, if a polarized source is necessary, a simple PBS will achieve this purpose. The second option is a filter stage. It could be useful if for instance, the detuning of the source must be accurately defined. In this case, an additional EDFA stage will be necessary to compensate the filter insertion loss. The filter can easily be realized with a Bragg grating, and slightly tuned by applying strain on it. We would like to stress that the proposed solution with an ASE source is very flexible and can easily be adapted to many cases.

Before implementing this solution in a cold atom experiment, it is important to estimate its noise level. The output of a 4 mW commercial erbium ASE source has been directly connected – without any optical filtering stage – to a 2 GHz band-pass photodetector. The photo-current is monitored with a high-speed real time 6 GHz oscilloscope. The sampling rate is set to $F_S$ = 20 GS/s. A set of 2 $10^7$ samples have been logged. The first one thousand samples of this raw signal are displayed Fig. 2(a). The mean value have been subtracted. As expected, this signal is noisy. It presents (for the entire set of data) a |~3 mV rms (root mean square) intensity noise level for a 168 mV mean value. This corresponds to a 57 SNR (Signal-to-Noise Ratio), or a ~2 % noise level. To take into account the fact that an atomic system will be sensitive to fluctuations typically below the MHz, the signal has been numerically filtered. Fig. 3 displays the SNR as a function of the cutoff frequency. From the right to the left, each point is obtained from the precedent one after a filtering stage, followed by a downsampling stage of the previous filtered signal.

Let's further detail this procedure. For a $F_S$ sampling frequency, $F_N = F_S / 2$ is the Nyquist frequency, the highest frequency that can be recovered. The 3 dB cutoff frequency $F_C$ of a low pass filter can be normalized with respect to $F_N$. $F_C$ ranges from 0 to 1. Without a downsampling stage, each point of Fig. 3 would require the design of a particular filter with each time a different cutoff frequency in order to process the whole set of logged data. Designing numerical filters for extremely low cutoff frequencies is delicate. For instance, $F_C = 10^{-3}$ requires a 4300 order Finite Impulse Response (FIR) filter. Beyond, our algorithm failed to design a $10^{-4}$ cutoff frequency. The technique in numerical signal processing is then the following. Let's consider a downsampling factor $D = 2$ (one sample is kept every $D$ samples), and the corresponding $F_C = 1 / D = 0.5$ filter. For this medium value, there is no particular problem, and the 30 order FIR filter that we have used will give accurate enough results. Once the filter has been applied, there are no significant informations anymore in the signal for frequencies ranging from 0.5 to 1. Downsampling the filtered signal by $D = 2$ will reduce the Nyquist frequency by the same factor without losing informations. Once the signal is downsampled, there is both a new Nyquist frequency which $1/D$ the previous one, and a new associated frequency space. The same filter might then been applied to the resulting signal, which can then downsampled, and so on... With this procedure there is no limit to the lower cutoff frequency (excepted of course the initial sample number). Furthermore, always the same well controlled filter is applied. The final obtained cutoff frequency is the cumulative product of the inverse of all the applied downsampling factors. Excepted for the first calculated point, all the point reported Fig. 3 are computed with $D = 2$ and $F_C = 0.5$. For the first point the parameters are $D = 10$ and $F_C = 10^{-1}$. Since for the raw data, $F_S = 20$ GS/s, this give a 1 GHz cutoff frequency for this first point, i.e. half the photodetector bandwidth.

According Fig. 3, for the ~1 MHz cutoff frequency, the SNR reaches ~2300, a 0.05 % noise level. This value is low enough for most applications (see the next section). The corresponding intensity signal is displayed Fig. 2(b). We have also checked that the choice of the filter is not critical. As an illustration, the filter might be applied twice at each stage, instead of once. This effectively doubles the filter order and its attenuation, i.e. 6 dB at $F_C$ instead of 3. As the filter is stronger, we expect a higher calculated SNR. However, the effect is weak. For instance, still at ~1 MHz, the SNR would be increased by only 1%, and by 7% at 60 kHz. The data reported Fig. 3 are thus good guidelines to estimate whether an experiment is feasible or not. An additional remark might be done: the SNR curve is well fitted with a $F^{-1/2}$ function. Indeed, the spectrum of an ASE source is flat on the GHz scale. Applying a $F_C = 1/2$ filter destroys half of the energy of the intensity fluctuation signal. The energy is proportional to the square of the amplitude of a signal, or to be more precise, to the square of its rms amplitude. Reducing the energy by a factor 2 reduces the rms value by $\sqrt{2}$, and as consequence, increases the SNR by $\sqrt{2}$. This explain the $\sim F^{-1/2}$ behavior of the SNR. It is also worthwhile to mention that if such a behavior is assumed to be valid for the high frequency range, the value SNR = 1 is reached for a ~6 THz cutoff frequency. Considering the simplicity of the reasoning, this is a rather good estimation of the ASE spectral width.

One should notice that if a lower noise level (or a wider bandwidth) is needed; it is possible to go further, as recently demonstrated for a Fiber Optical Parametric Amplifier (FOPA) [15]. Thanks to Kerr effect, the intensity noise of an ASE source can be transferred as phase noise to a low intensity noise DFB laser diode; at the cost however, of a more complex setup. The next section studies the effect of the estimated noise level for a BEC cooling experiment.

## 3. Noise level and heating rate

A potential problem of the proposed scheme is the heating of atoms (and the associated losses) induced by the intensity fluctuations of the beams. Heating in a dipole trap has been modeled in [12] by considering an atom trapped in a fluctuating harmonic potential $V(t) = k(t)x^2/2$:

$$H = \frac{p^2}{2M} + \frac{1}{2}k(t)x^2 \qquad (1)$$

with $k(t) = k_0(1 + \varepsilon(t))$, $\varepsilon(t)$ being a stochastic function with a coherence time $\tau_0$. As $k \propto d^2V/dx^2 \propto I$, $\varepsilon(t) \ll 1$ corresponds to the fractional fluctuation of the radiation intensity $I$. Writing

$$H = H_0 + \frac{1}{2}k_0\varepsilon(t)x^2 \qquad (2)$$

reduces the problem to the calculus of a quadratic perturbation of a harmonic oscillator. As shown in [12], this perturbation only couples eigenstates $|n\rangle$ (of eigenenergy $E_n = n\hbar\sqrt{k_0/M} = n\hbar\omega_0$) and $|m\rangle$ such that $n = m \pm 2$. Hence, only the spectral components of the noise around integer multiples of $2\omega_0$ contribute to the heating process.

We shall consider here radiation of noise spectrum $S(\omega)$, of bandwidth $\tau_0^{-1}$, and with *constant* total noise power $\int S(\omega)d\omega = S_0$; for the sake of definiteness, we will take a spectrum of width $\tau_0^{-1}$ and constant amplitude $S_0\tau_0$. As $\tau_0$ diminishes, the amplitude of the noise diminishes, but the width of the spectrum increases, including more harmonics of $2\omega_0$. These new harmonics included in the spectrum tend to increase heating; on the other hand they contribute only in higher orders of perturbation theory. It is thus reasonable to expect that using a large spectrum of small amplitudes (with constant total power) lead to an overall *reduction* of heating and losses. In order to verify that quantitatively, we performed a numerical simulation of the dynamics of the heating process.

Writing the general solution of the Schrödinger equation associated with Eq. (2) in the form $|\psi(t)\rangle = \sum_n a_n(t)e^{-i\omega_0 t}|n\rangle$, one founds, after a simple algebra, a system of coupled ordinary differential equations for the coefficients $a_n(t)$, which can be easily solved numerically by e.g. a Runge-Kutta routine. For any time $t$ one can then calculate the average system energy $E = \hbar\omega_0 \sum_n n|a_n(t)|^2$ and deduce from that the heating rate $\dot{E} = dE/dt$. For high enough excitation, the optical potential departs from the harmonic approximation. However, for a typical laser beam with a Gaussian transverse profile, atoms having a high enough energy will escape from the potential. In most cases, one is interested only in low energy atoms (which are in the harmonic range), so, in our simulation, we simply considered that atoms arriving at a high enough level $n = n_{max}$ (with typically $n_{max} = 12$ in the simulations) are lost. For a trap frequency $\omega_0/2\pi = 3$ kHz (see below), for example, $n_{max} = 12$ corresponds to atoms at a temperature ~1.8 μK. In e.g. evaporative cooling in a condensation experiment, atoms are expelled from the trap well below such temperatures. Thus, this simple procedure allows us to obtain a reasonable evaluation of the loss rate on the process.

Fig. 4 displays typical curves of heating and losses rates in a log-log plot. It clearly shows that the heath rate decreases significantly for $\tau_0 < (2\omega_0)^{-1}$. The loss rate is small even for higher values of $\tau_0$, becoming important only when the amplitude of the noise is high enough that higher order couplings come into play. This simple simulation thus confirms quantitatively that using a large bandwidth optical source with a low spectral density has very favorable properties with respect to heating and losses of atoms.

For a chosen filter cutoff frequency, and from Fig. 4, one may determine the corresponding heating rate to determine whether this value is acceptable for a particular experiment. Let us now consider a real application case: a 1560 nm optical-dipole trap for $^{87}$Rb BEC production made by a tightly confining beam, as used in [6]. The heating rate due to intensity fluctuations of the beam can be estimated, according to our model (Fig 4.). First, we must pay attention to the waist value and the optical power. An initial collision rate (~ $10^4$ s$^{-1}$ for 3 millions of $^{87}$Rb atoms) and an initial potential depth U (U/k$_B$ ~ 900 μK) are required for an efficient evaporative cooling process towards Bose-Einstein condensation. We thus estimate the heating rate in the ASE-based optical trap made by one focused light beam: a 28 micrometer waist and a 6 Watt power give transverse and longitudinal trapping frequencies of respectively 3.35 kHz and 42 Hz. In the case of an ASE-based optical-dipole trap with a ~1 MHz bandwidth as described above, we obtain a heating rate for each frequency which is well below $10^{-10}$ $\hbar\omega_0^2$ (and as a consequence out of the diagram Fig 4.). Since $10^{-10}$ $\hbar\omega_0^2$ corresponds to 3 pK/s (pico-Kelvin per second), the heating rate would thus be completely negligible for a real experimental realization compared to the potential depth and to others losses mechanism such as laser-beam-pointing noise, collisions with residual gas... For instance, the heating rate induced by spontaneous photon scattering is 30 nK/s at the trap center, i.e. a rate larger by four orders of magnitude. Furthermore, the chosen cutoff frequency is well above the trapping frequencies giving a high level of confidence to this result.

One can also consider a single-beam optical trap at the same wavelength dedicated to single atom manipulation. Optical trap parameters, such as waist and optical power, are chosen to give a sufficient potential depth for single-atom manipulation and a tiny trap volume where only one atom can be loaded [16]. With a 3 micrometers waist and a 100 mW optical power, the corresponding frequencies for each direction are respectively: 37.6 kHz, 37.6 kHz and 4.4 kHz. Similarly, the heating rate due to intensity fluctuations is negligible with a value below $10^{-10}\ \hbar\omega_0^2$, i.e. 300 pK/s. Once again, this is a negligible value compared with the heating rate related to the spontaneous scattering of photons for the same trap parameters, i.e. 43 nK/s.

## 4. Conclusion

In summary, we have proposed an optical-dipole trap technique based on an Amplified Spontaneous Emission source. This novel type of application for BEC production can be easily implemented in a typical cold-atom apparatus. The main advantage of this trapping scheme is its low coherence length. It should easily allow the creation of multi-beam geometry configurations in order to optimize both loading and evaporation stages. Furthermore this source as a seed is very stable and flexible. It has all the advantages of the fibered optical telecommunications technology and can simply be adapted to many particular experimental requirements. Although this source is naturally very noisy, the heating rate estimated for real world potential applications is found to be completely negligible in the range of the pK/s to a few hundred. Furthermore, the heating rate related to the light intensity fluctuations of our filtered ASE source is much lower than the heating rate related to spontaneous photon scattering by a few orders of magnitude for BEC all-optical production and single-atom optical trap. Finally, this method can be extended to numerous situations where optical trapping is needed: cold molecules, single atom experiments, nanoparticles manipulations, and optical tweezers for biology.

**Acknowledgements**

Jean-François Clément is supported by the project "MICPAF", ANR-07-BLAN-0137-01 and by a CNRS grant. This work is partially being performed under the auspices of *Région Nord-Pas-de-Calais* and *European Union* (FEDER).

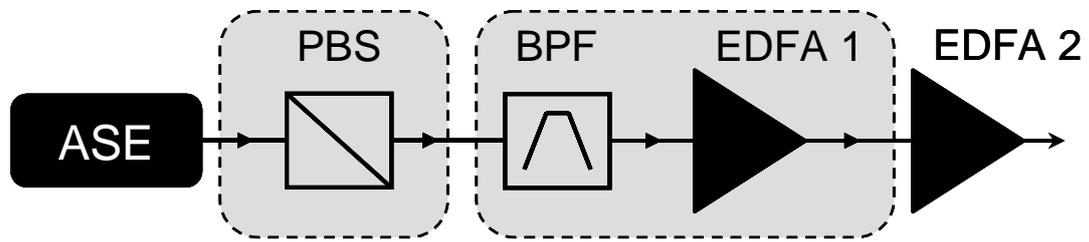

Fig. 1. Schematic view of the proposed device. ASE: erbium amplified spontaneous emission source. PBS: Polarization beam splitter. BPF: band-pass filter. EDFA 1: low power amplifier. EDFA 2: High power erbium doped fiber amplifier. The grey insets show optional components.

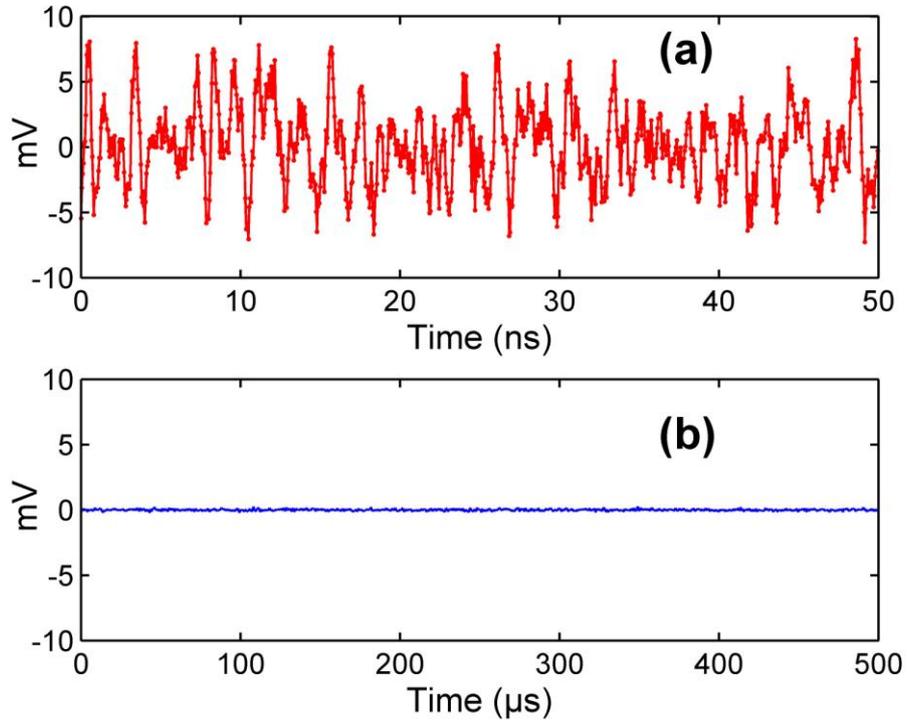

Fig. 2. **(a)** ASE Intensity signal trace as a time function of time (1000 samples). The mean value (~168 mV) has been subtracted. The acquisitioned is performed with a 2 GHz photodetector. Noise level: ~2 %. **(b)** Filtered and downsampled signal with a ~1 MHz cutoff frequency at ~3 dB (1000 samples). Noise level: ~0.05 %.

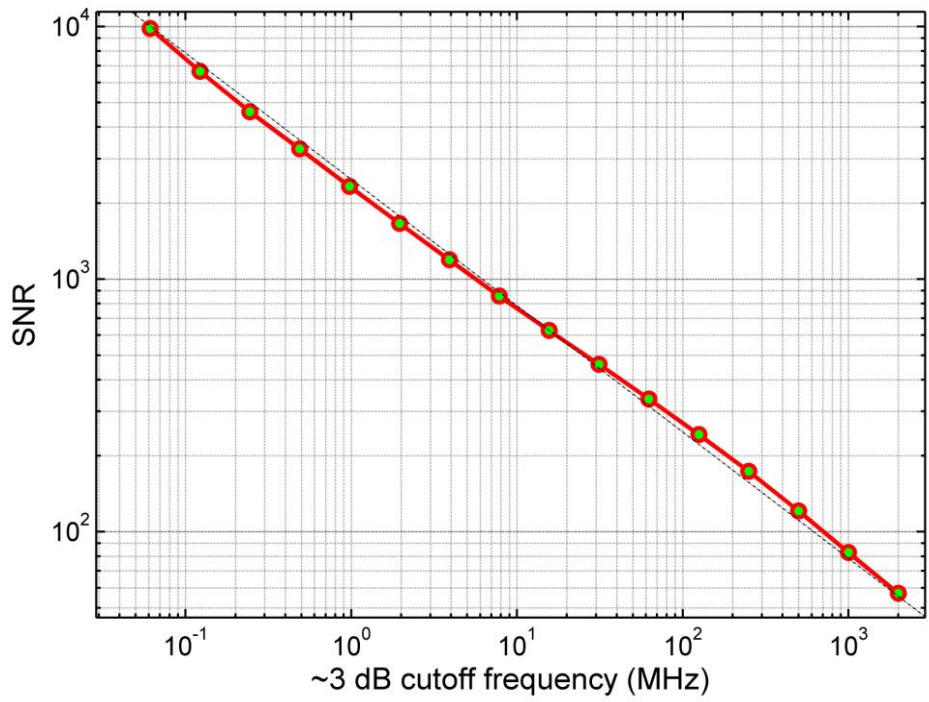

Fig. 3. Red solid line: estimation of the signal-to-noise ratio (SNR) of the source intensity as a function of the cutoff frequency. Black dashed line: fit of the SNR curve with a $F^{-1/2}$ function.

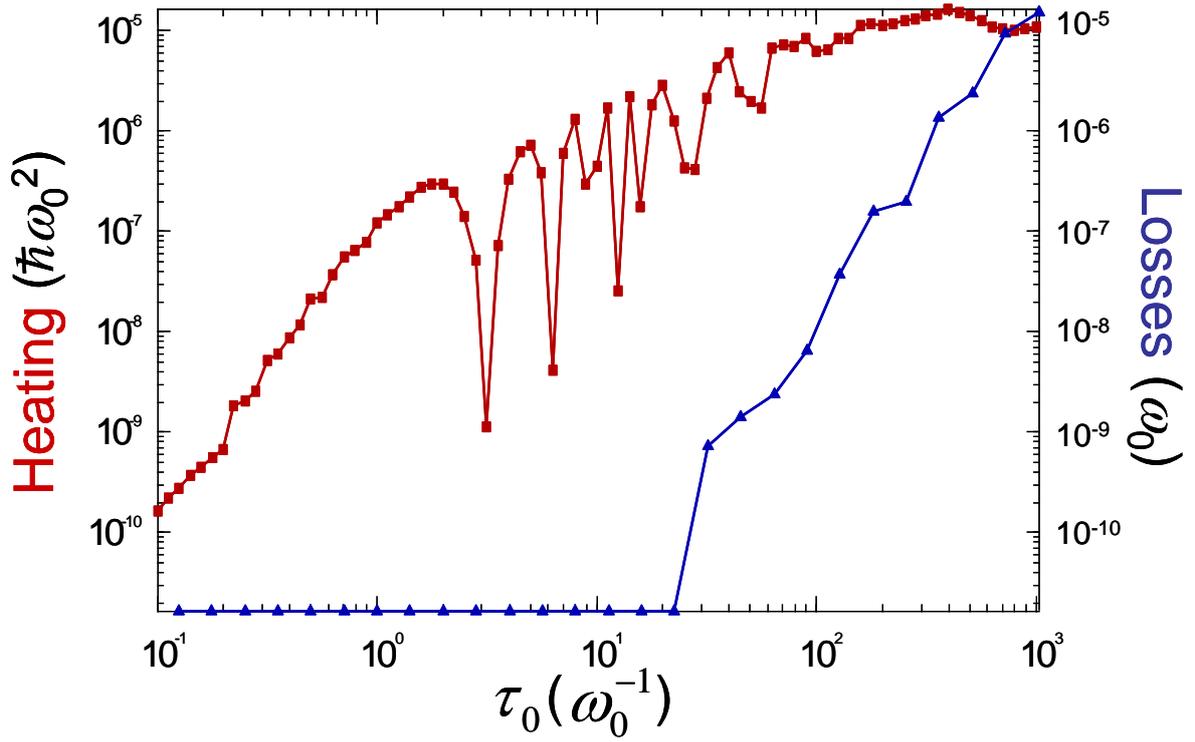

Fig. 4. Numerical calculation of the heating rate $\dot{E}$ (red squares) and losses (blue triangles). The effect of the successive resonances is clearly visible in the heating rate for coherence times comparable to the trap frequency $\omega_0$. The horizontal values of the loss rate mean that it is to low to be calculated.